%% file: main.tex
\documentclass[10pt,conference]{IEEEtran}
\IEEEoverridecommandlockouts
\usepackage{cite}
\usepackage{amsmath,amssymb,amsfonts}
\usepackage{graphicx}
\usepackage{textcomp}

\usepackage{algorithm}
\usepackage{algpseudocode}
\usepackage{multirow}

\usepackage{xcolor}
\def\BibTeX{{\rm B\kern-.05em{\sc i\kern-.025em b}\kern-.08em
    T\kern-.1667em\lower.7ex\hbox{E}\kern-.125emX}}

\newcommand{\tool}{\texttt{NavAI}}
    
\begin{document}

\title{NavAI: A Generalizable LLM Framework for Navigation Tasks in Virtual Reality Environments
}


\author{\IEEEauthorblockN{Xue Qin}
\IEEEauthorblockA{\textit{Department of Computing Sciences} \\
\textit{Villanova University}\\
Villanova, U.S. \\
xue.qin@villanova.edu}
\and
\IEEEauthorblockN{Matthew DiGiovanni}
\IEEEauthorblockA{\textit{Department of Computing Sciences} \\
\textit{Villanova University}\\
Villanova, U.S. \\
mdigio02@villanova.edu}
}

\maketitle

\begin{abstract}
Navigation is one of the fundamental tasks for automated exploration in Virtual Reality (VR). Existing technologies primarily focus on path optimization in 360-degree image datasets and 3D simulators, which cannot be directly applied to immersive VR environments. To address this gap, we present \tool{}, a generalizable large language model (LLM)-based navigation framework that supports both basic actions and complex goal-directed tasks across diverse VR applications. We evaluate \tool{} in three distinct VR environments through goal-oriented and exploratory tasks. Results show that it achieves high accuracy, with an 89\% success rate in goal-oriented tasks. Our analysis also highlights current limitations of relying entirely on LLMs, particularly in scenarios that require dynamic goal assessment. Finally, we discuss the limitations observed during the experiments and offer insights for future research directions.
\end{abstract}

\begin{IEEEkeywords}
Virtual Reality, Large Language Model, Navigation
\end{IEEEkeywords}

\input{content/intro}

\input{content/framework}
\input{content/evaluation}
\input{content/discussion}
\input{content/conclusion}

\bibliographystyle{IEEEtran}
\bibliography{reference}

\end{document}

%% file: content/intro.tex
\section{Introduction}

According to industry forecasts, consumer VR and AR shipments are expected to grow from 17.81 million units in 2023 to 30.88 million by 2026~\cite{FortuneBusinessInsights2025VRMarket}.
With its ability to simulate physical environments, VR is now widely used in applications such as real estate tours, science education, and manufacturing training.
To ensure quality at scale, researchers and developers have proposed and applied automated testing~\cite{Dhia2023issta} in VR applications, 
focusing on areas such as exploration testing ~\cite{Wang2022VRTest, qin2024utilizinggenai}, unit testing~\cite{Gil2020Youkai}, and privacy enhancement~\cite{kim2025autovrautomateduiexploration}.

Among all these automated testing tasks, a successful navigation serves as one of the fundamental starting points for the exploration to help the user interact with the environment effectively. Navigating in a VR space—such as an unknown city or dense forest—can be challenging without guidance, and a map alone is insufficient for addressing spatial navigation, user orientation, and engagement. 
The current state-of-the-art solution for navigation is called Vision Language Navigation (VLN), which requires the agent to follow human language instructions to navigate in previously unseen environments. 
Types of VLN include Goal-oriented~\cite{REVERIE2020CVPR} and Route-oriented~\cite{Anderson2018VLNCVPR, ku-etal-2020-room}, and the user may interact with the VLN in single turn or multiple turns~\cite{Nguyen2019cvpr}.

Most discussed VLN techniques have been developed and tested either on real-world environments represented through 360 virtual tours or within traditional 3D simulators. However, these environments differ significantly from fully immersive VR settings.
Existing navigation support for VR is typically limited to either pre-defined guides embedded in pre-programmed environments~\cite{Collins2024aiguideassets} or built-in AI features tailored to specific VR applications~\cite{konenkov2024vrgptvisuallanguagemodel}.
As the VR industry continues to grow, many new and complex environments will be created, and new engagements will be designed.
There is a pressing need for robust and scalable navigation solutions that can generalize across both current and emerging VR applications.

In this paper, we propose a large language model (LLM)-based navigation solution framework, named \tool{}, that can be applied across diverse VR environments.
The framework leverages the broad knowledge of LLMs to interpret the virtual world using screenshots captured from the VR application. 
Specifically, \tool{} first employs the Comprehensive Interpreter component to analyze the visual scene and generate both visual interpretations and textual descriptions.
Then, the Multi-agent Decision Voter parses the user's navigation request and categorizes it into one of the newly defined navigation types. This voter also helps decide whether the intended goal has been achieved.
If the navigation task is still in progress, the Decision-to-Control Mapping matches the intent into corresponding control functions and executes navigation actions.
\tool{} currently supports both basic action commands and complex goal-directed navigation tasks.

We evaluate the effectiveness of \tool{} across three diverse VR environments and assign different navigation tasks of goal-driven and exploratory scenarios. 
The framework shows consistent performance on basic action tasks such as ``move forward'' and ``turn left'', successfully executing all 21 basic motion commands with minimal overhead (average 0.74s).
\tool{} also achieves an 89\% success rate (16 out of 18) in direct goal-oriented tasks. However, efficiency varies across environments, with the interpretation component and decision-making voter introducing notable latency, particularly in tasks requiring flexible stopping conditions like exploratory scanning.
Finally, we conduct an in-depth discussion on the usability of \tool{} and address the observed limitations by proposing potential future solutions.

The contributions of this paper are as follows:
\begin{itemize}
    \item We designed a generalizable framework that can perform navigation tasks in any VR environment by bridging user requests with interpretations and decisions made by the Large Language Models (LLMs).
    \item We conduct three sets of experiments covering different task types across diverse VR environments. Results demonstrate that \tool{} achieves high success rates and reasonable efficiency across scenarios.
    \item We analyze the usability of LLM-based navigation, highlighting current limitations and proposing future directions to improve performance and real-time applicability. 
\end{itemize}

%% file: content/framework.tex
\section{Framework}

\begin{figure}[h]
  \centering
  \includegraphics[width=\linewidth]{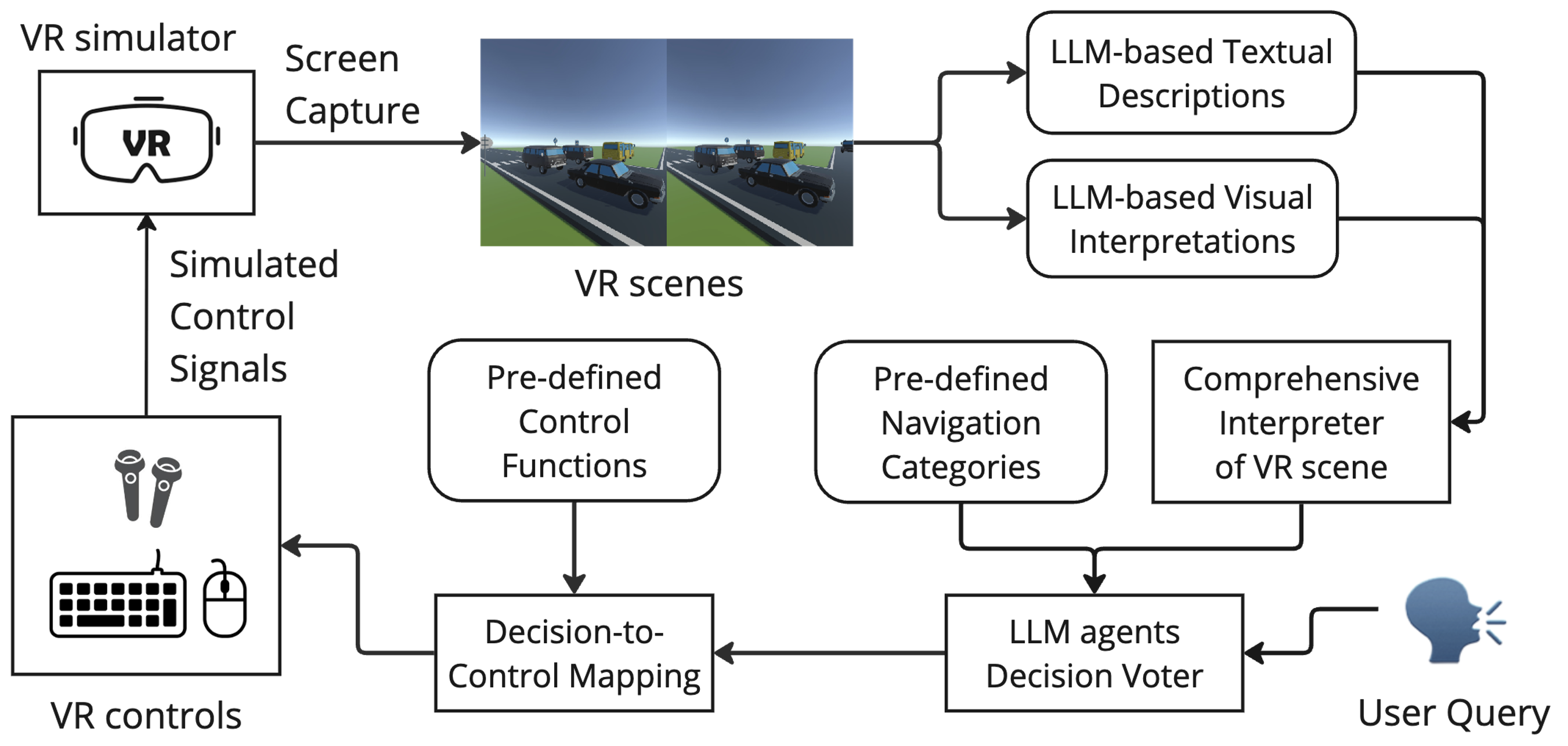}
  \caption{\tool{} Framework overview}
  \label{fig:overview}
\end{figure}

In this section, we introduce the \tool{} framework, which automatically captures and processes stereoscopic images from a VR application, analyzes the user’s query, and then sends the corresponding simulated control signals.
As illustrated in Figure~\ref{fig:overview}, the \tool{} consists of the following core components for navigation automation:
1) Comprehensive Interpreter – leverages a Large Language Model to generate both textual descriptions and visual interpretations of VR scenes.
2) Pre-defined Navigation Categories – a dataset covering common navigation types and subtypes to guide user query classification.
3) Decision Voter – aggregates the decisions of multiple LLM agents to categorize the intended navigation action.
4) Decision-to-Control Mapping – matches categorized queries to corresponding control functions and executes them on the VR simulator.
Other supporting components, such as the screen capture module and user query receiver, are excluded from this discussion as they are not central to the navigation automation pipeline.

\subsection{Comprehensive Interpreter}

Since all the VR scenes are never trained and seen before, the main goal of the comprehensive interpreter is to have the Large Language Model generate a basic understanding of the VR environment so that the user's queries can proceed easily with context.
Cognitive theories—particularly dual-coding theory~\cite{Paivio1990DualCoding}—suggest that the human brain encodes both visual and verbal information when processing an object. Following this perspective, we assign two types of tasks to the LLM when analyzing VR screenshots (representing the user’s field of view, FOV): visual interpretation and textual description.
In particular, for text description, we adopt the chain-of-thought design from the existing VR study~\cite{qi2025harnessinglargelanguagemodel} to collect the observed objects' names and their feature descriptions. 
For visual interpretation, we automatically overlay a grid on top of the screenshots and ask the LLM to provide coordinate information.
We do not segment and crop every single object from the screen due to time cost considerations.
We also consider coordinates to be more useful, as they can indirectly represent the spatial relationships between object positions and the user’s standing or observation point.
At the end, we aggregated both types of information as context for the decision voter to decide which navigation action should be performed.

\subsection{Pre-defined Navigation Categories}
\begin{figure}
  \centering
  \includegraphics[width=0.9\linewidth]{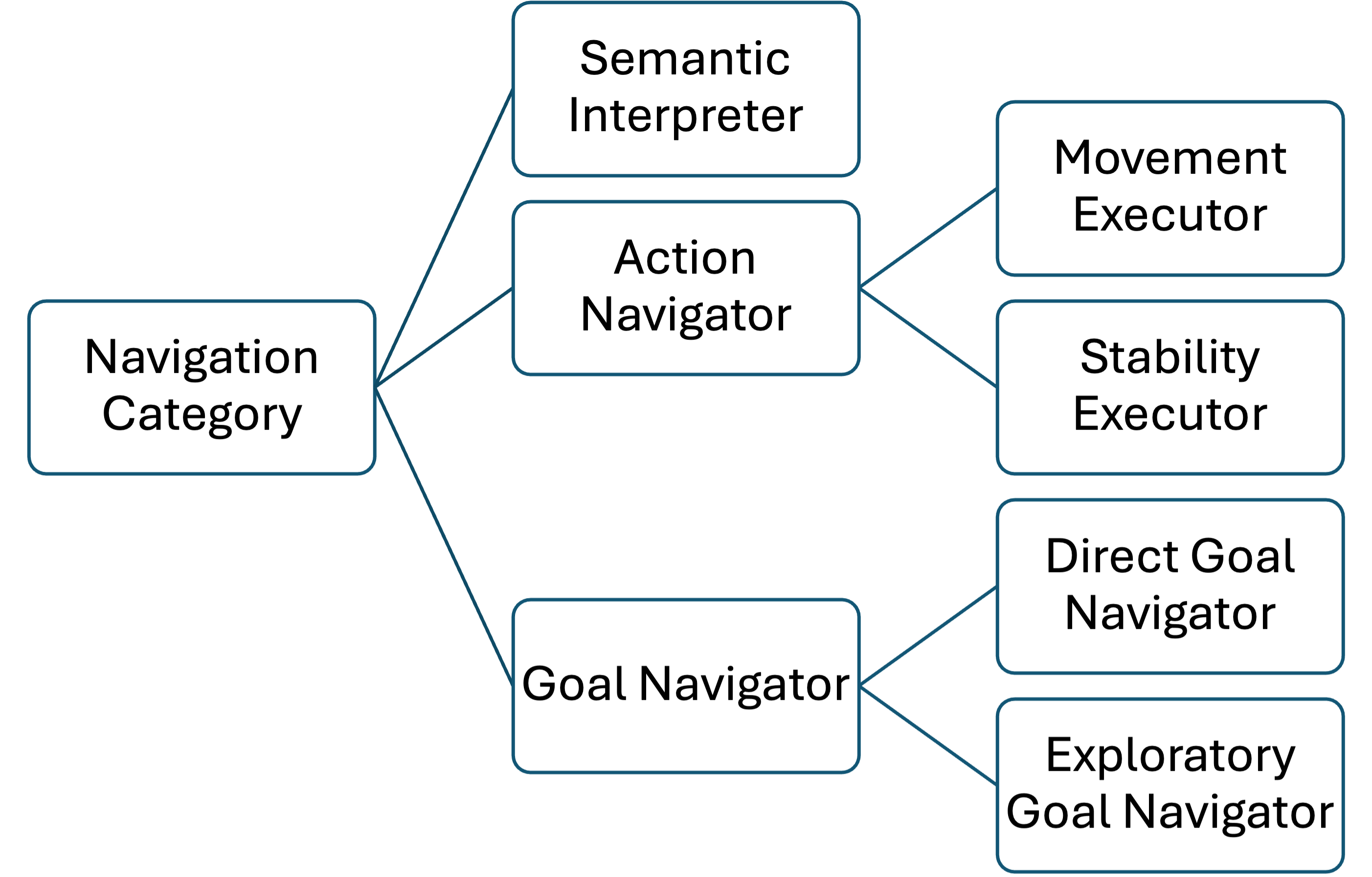}
  \caption{Navigation Categories for Exploration}
  \label{fig:nav_cat}
\end{figure}

An existing taxonomy of navigation tasks was built from human instructions in a survey conducted by Wu et al.~\cite{wu2022visionlanguagenavigationsurveytaxonomy}. The survey examined various navigation tasks in 3D simulators that embed embodied AI, where the built-in AI agents act and interact with the environment through extensive training.
In this paper, we focus on navigation tasks that support automated exploration, with particular attention to visiting on-site objects that are visible within the FOV or present in the VR scenes. Figure~\ref{fig:nav_cat} presents our refined taxonomy, which consists of three major types.

\begin{itemize}
    \item The first category is the \textbf{Semantic Interpreter}, which handles non-navigation queries such as ``Where am I?'' These tasks can be resolved without sending control signals or triggering movement.
    \item The second category is the \textbf{Action Navigator}, where the user explicitly requests a basic action, such as ``move forward” or “turn left.'' This category is divided into two subtypes: Movement Executor and Stability Executor, as they invoke different groups of control functions.
    \item The third category is the \textbf{Goal Navigator}, where the user specifies a navigation request that links to a target object visible in the FOV, such as ``I want to go to the yellow bus.'' This type is more complex since the user does not specify the particular actions. This task typically requires a sequence of actions and will be completed in multiple turns. We further divide this into two subtypes: Direct Goal Navigator, which refers to goals visible in the current FOV, and Exploratory Goal Navigator, which requires scanning or searching for the target before executing the navigation.
\end{itemize}

\subsection{Decision Voter}

The Decision Voter aggregates the outputs from the Comprehensive Interpreter and the Pre-defined Navigation Categories, and determines the appropriate response for the navigation system based on the user’s query. 
Specifically, it first classifies the query into a navigation category, then evaluates whether the user’s requirement has been satisfied or the task goal has been achieved. If the goal is not met, the voter proceeds to the action decision stage to guide the next step.

\begin{algorithm}
\caption{Decision Voter for Navigation}\label{alg:decision_voter}
    \begin{algorithmic}[1]
    \Procedure{DecisionVoter}{$query$}
    \State $ctx \gets \textsc{GetSceneContext}()$
    \State $cat \gets \textsc{ClassifyCategory}(query, ctx)$
    \If{$cat = \textsc{SEMANTIC\_INTERPRETER}$}
        \State \Return $\textsc{SEMANTIC\_INTERPRETER}$
    \ElsIf{$cat = \textsc{ACTION\_NAVIGATOR}$}
        \State \Return $\textsc{ACTION\_NAVIGATOR}$
    \Else
        \State $v \gets \textsc{VoteAgents}(query, ctx)$
        \If{$v = \textsc{REACHED}$}
            \State \textsc{StopNav}()
            \State \Return \textsc{GOAL\_REACHED}
        \Else
            \State \Return \textsc{GOAL\_PROGRESS}
        \EndIf
    \EndIf
    \EndProcedure
    \end{algorithmic}
\end{algorithm}

The detailed algorithm is shown in Algorithm~\ref{alg:decision_voter}.
The \textsc{DecisionVoter} procedure receives a user query in natural language and determines the appropriate navigation action. 
First, the VR scene context is obtained through the method $GetSceneContext()$ 
at line 2. 
The query is then classified into one of three categories at line 3.
As illustrated from line 4 to line 7, 
the procedure will return either \textsc{SEMANTIC\_INTERPRETER} or \textsc{ACTION\_NAVIGATOR} if the category type matches the first two cases.
For goal-based navigation, multiple LLM agents vote on whether the intended goal, as referred to in the user query, has been reached, given the scene context. 
If the goal is confirmed, navigation is stopped and the system returns \textsc{GOAL\_REACHED}. 
Otherwise, the system returns \textsc{GOAL\_PROGRESS}, and the following Decision-to-Control Mapping component will decide and execute the next control action.

\subsection{Decision-to-Control Mapping}
After determining the goal status, the final Decision-to-Control Mapping component identifies and executes the appropriate control function from a set of pre-defined basic actions. 
In practice, this mapping connects the user’s navigation intent with low-level control commands in the VR environment. 
We found that OpenAI’s function calling feature aligns well with this need, as it allows us to expose the action functions as JSON objects and let the model generate the most suitable invocations, including both parameters and values. 
This mechanism provides a clean separation between high-level natural language interpretation and concrete execution. 
The complete set of pre-defined basic actions, categorized into \emph{Movement} and \emph{Stability}, is shown in Table~\ref{tab:basic_actions}. 
Movement actions control positional changes, such as moving forward, while stability actions keep the user in place but adjust orientation or view, including a 360-degree scene scan for comprehensive exploration.

\begin{table}[h]
\centering
\caption{Pre-defined Basic Navigation Actions}
\label{tab:basic_actions}
\begin{tabular}{|l|l|}
\hline
\textbf{Type} & \textbf{Action Functions} \\ \hline
\multirow{4}{*}{Movement} 
    & \texttt{move\_forward(duration)} \\
    & \texttt{move\_left(duration)} \\
    & \texttt{move\_right(duration)} \\ \hline
\multirow{5}{*}{Stability} 
    & \texttt{in\_place\_rotate\_to\_left(duration)} \\
    & \texttt{in\_place\_rotate\_to\_right(duration)} \\
    & \texttt{look\_up(duration)} \\
    & \texttt{look\_down(duration)} \\
    & \texttt{scan\_360()} \\ \hline
\end{tabular}
\end{table}

%% file: content/evaluation.tex
\section{Evaluation}

In this section, we evaluate the effectiveness of the \tool{} framework. Specifically, we measure the success rates of different task types across multiple environments and analyze execution times to assess overall usability. We exclude the Semantic Interpreter tasks from evaluation, as they are primarily concerned with the image recognition accuracy of the AI rather than navigation performance.

\subsection{Dataset and Setup}

We selected three different Unity VR assets as testing environments. 
The first is a \textit{Highway} scene, which includes multiple cars and an intersection. 
The second is a \textit{Ship} scene, featuring an open deck within a narrow space. 
The third is a \textit{Country House}, an indoor environment containing multiple rooms and objects. 
To run these assets, we use the Unity Simulator to simulate the VR environments without requiring a physical headset. 
The simulator is well-suited for functional testing, which aligns with our focus on evaluating the effectiveness of navigation decision algorithms. 
For the \textit{Comprehensive Interpreter} and the \textit{Decision-to-Signal Mapping} components, we employ GPT-4o~\cite{openai2024gpt4o}, following the study by Qi et al.~\cite{qi2025harnessinglargelanguagemodel}. 
In the \textit{Decision Voter} component, we integrate three different LLMs for major voting strategies: GPT-4o~\cite{openai2024gpt4o}, Grok-2~\cite{xai2024grok2}, and Gemini-2.5-Flash~\cite{google2025gemini}.

\subsection{Effectiveness of Action Navigator Task}

Table~\ref{tab:action_navigator} reports the execution durations of \tool{} when the user’s query falls into the Action Navigator category across three VR environments. Each action was designed to last 2 seconds, and all were executed successfully. On average, the measured overhead across all actions and environments is approximately 0.74 seconds. This overhead reflects the combined time cost of GPT-4o in mapping the request to signal functions, the execution of the defined functions within \tool{}, and the simulation and rendering costs introduced by the Unity environment. Overall, the results indicate that \tool{} can process and execute basic action queries consistently and efficiently across diverse VR contexts.

\begin{table}[ht]
\centering
\caption{Execution durations (in seconds) and average overhead of Action Navigator queries across three VR environments.}
\label{tab:action_navigator}
\begin{tabular}{|l|c|c|c|c|}
\hline
\textbf{Action} & \textbf{Highway} & \textbf{Country House} & \textbf{Ship} & \textbf{ Overhead} \\ \hline
Move Forward      & 2.68 & 2.74 & 2.72 & 0.71 \\ \hline
Move Left         & 2.72 & 2.62 & 2.61 & 0.65 \\ \hline
Move Right        & 2.60 & 3.49 & 2.75 & 1.28 \\ \hline
Rotate Left  & 2.59 & 2.82 & 2.63 & 0.68 \\ \hline
Rotate Right & 2.45 & 2.80 & 2.65 & 0.63 \\ \hline
Look Up      & 2.76 & 2.95 & 2.70 & 0.80 \\ \hline
Look Down    & 2.60 & 2.91 & 2.67 & 0.73 \\ \hline
\textbf{Overall Average} & 2.63 & 2.90 & 2.67 & \textbf{0.74} \\ \hline
\end{tabular}
\end{table}

\begin{table*}[t]
\centering
\caption{Evaluation results for Direct Goal-oriented navigation in the Highway environment.}
\label{tab:highway_eval}
\begin{tabular}{|c|c|c|c|p{2.2cm}|p{2.2cm}|c|c|}
\hline
\multirow{2}{*}{\textbf{Attempt}} & 
\multirow{2}{*}{\textbf{Success}} & 
\multirow{2}{*}{\textbf{\# Turns}} & 
\multicolumn{1}{c|}{\textbf{Decision Voter}} & 
\multicolumn{2}{c|}{\textbf{Comprehensive Interpreter}} & 
\multicolumn{1}{c|}{\textbf{Action}} & 
\multicolumn{1}{c|}{\textbf{Total/Turn}} \\ \cline{5-6}
 &  &  & \textbf{(sec.)} & \centering \textbf{Visual (sec.)} & \centering \textbf{Textual (sec.)} & \textbf{(sec.)} & \textbf{(sec.)} \\ \hline
1 & Yes & 9  & 8.35 & \centering 13.02 & \centering 15.00 & 3.60 & 40.51 \\ \hline
2 & Yes & 11 & 10.33 & \centering 14.60 & \centering 16.13 & 4.52 & 46.23 \\ \hline
3 & Yes & 10 & 11.61 & \centering 13.62 & \centering 16.18 & 5.31 & 47.32 \\ \hline
4 & Yes & 7  & 13.48 & \centering 19.78 & \centering 17.90 & 3.99 & 55.88 \\ \hline
5 & Yes & 7  & 8.26  & \centering 15.39 & \centering 16.80 & 3.82 & 44.95 \\ \hline
6 & Yes & 13 & 9.09  & \centering 17.97 & \centering 17.92 & 4.39 & 50.09 \\ \hline
\textbf{Average} & - & \textbf{9.5} & \textbf{10.19} & \centering \textbf{15.73} & \centering \textbf{16.66} & \textbf{4.27} & \textbf{47.50} \\ \hline
\end{tabular}
\end{table*}

\begin{table*}[t]
\centering
\caption{Evaluation results for Direct Goal-oriented navigation in the Country House environment.}
\label{tab:countryhouse_eval}
\begin{tabular}{|c|c|c|c|p{2.2cm}|p{2.2cm}|c|c|}
\hline
\multirow{2}{*}{\textbf{Attempt}} & 
\multirow{2}{*}{\textbf{Success}} & 
\multirow{2}{*}{\textbf{\# Turns}} & 
\multicolumn{1}{c|}{\textbf{Decision Voter}} & 
\multicolumn{2}{c|}{\textbf{Comprehensive Interpreter}} & 
\multicolumn{1}{c|}{\textbf{Action}} & 
\multicolumn{1}{c|}{\textbf{Total/Turn}} \\ \cline{5-6}
 &  &  & \textbf{(sec.)} & \centering \textbf{Visual (sec.)} & \centering \textbf{Textual (sec.)} & \textbf{(sec.)} & \textbf{(sec.)} \\ \hline
1 & Yes      & 5  & 6.38  & \centering 17.37 & \centering 17.32 & 4.75 & 46.65 \\ \hline
2 & Yes      & 5  & 5.39  & \centering 11.23 & \centering 13.58 & 3.49 & 34.17 \\ \hline
3 & Yes      & 5  & 5.53  & \centering 10.94 & \centering 15.38 & 3.08 & 35.37 \\ \hline
4 & Yes      & 5  & 19.55 & \centering 17.38 & \centering 17.38 & 5.32 & 63.43 \\ \hline
5 & Yes      & 4  & 7.57  & \centering 7.12  & \centering 20.07 & 7.49 & 42.59 \\ \hline
6 & No       & 19 & 6.49  & \centering 14.63 & \centering 15.92 & 6.98 & 44.77 \\ \hline
\textbf{Average} & - & \textbf{7.2} & \textbf{8.82} & \centering \textbf{13.11} & \centering \textbf{16.61} & \textbf{5.18} & \textbf{44.50} \\ \hline
\end{tabular}
\end{table*}

\begin{table*}[t]
\centering
\caption{Evaluation results for Direct Goal-oriented navigation in the Ship environment.}
\label{tab:ship_eval}
\begin{tabular}{|c|c|c|c|p{2.2cm}|p{2.2cm}|c|c|}
\hline
\multirow{2}{*}{\textbf{Attempt}} & 
\multirow{2}{*}{\textbf{Success}} & 
\multirow{2}{*}{\textbf{\# Turns}} & 
\multicolumn{1}{c|}{\textbf{Decision Voter}} & 
\multicolumn{2}{c|}{\textbf{Comprehensive Interpreter}} & 
\multicolumn{1}{c|}{\textbf{Action}} & 
\multicolumn{1}{c|}{\textbf{Total/Turn}} \\ \cline{5-6}
 &  &  & \textbf{(sec.)} & \centering \textbf{Visual (sec.)} & \centering \textbf{Textual (sec.)} & \textbf{(sec.)} & \textbf{(sec.)} \\ \hline
1 & Yes & 17 & 7.35 & \centering 29.53 & \centering 18.42 & 5.33 & 62.00 \\ \hline
2 & No  &  4 & 5.66 & \centering 20.63 & \centering 18.31 & 3.62 & 49.30 \\ \hline
3 & Yes &  6 & 6.34 & \centering 31.51 & \centering 20.27 & 3.37 & 63.14 \\ \hline
4 & Yes &  4 & 5.18 & \centering 14.13 & \centering 15.58 & 4.47 & 40.02 \\ \hline
5 & Yes &  7 & 6.45 & \centering 27.14 & \centering 19.30 & 4.32 & 58.72 \\ \hline
6 & Yes &  6 & 6.03 & \centering 33.92 & \centering 18.75 & 4.33 & 64.40 \\ \hline
\textbf{Average} & - & \textbf{7.3} & \textbf{6.17} & \centering \textbf{26.48} & \centering \textbf{18.77} & \textbf{4.24} & \textbf{56.26} \\ \hline
\end{tabular}
\end{table*}

\subsection{Effectiveness of Direct Goal Navigator Task}
We further evaluate the effectiveness of \tool{} on Direct Goal-oriented navigation tasks, focusing on both accuracy (success rate) and efficiency (execution time). Representative tasks were tested across three VR environments. In the Highway scene, the task was ``\textit{Get to the back of the yellow bus and avoid hitting other cars, going around them.}'' In the Country House, the task was ``\textit{Walk through the doorway on the left and enter the bedroom.}'' In the Ship environment, the task was ``\textit{Walk over to the cannon on your right.}''
Tables~\ref{tab:highway_eval},\ref{tab:countryhouse_eval},\ref{tab:ship_eval} present the execution outcomes across multiple attempts in each environment, along with detailed efficiency measurements in terms of time per navigation turn.

Across all environments, 16 out of 18 attempts were ultimately successful, indicating that \tool{} can reliably reach designated goals. However, the efficiency analysis highlights that navigation incurs notable time costs. On average across the three testing environments, the execution time per turn is approximately 49.4 seconds, with an average of 8 turns per attempt. This means that completing a typical goal-oriented navigation task can take up to 395 seconds in total. 
Within each turn, the Comprehensive Interpreter dominates with about 72\% (35.6 seconds), consisting of Visual interpretation (19.8 seconds, 40\%) and Textual description (15.8 seconds, 32\%). The Decision Voter contributes around 17\% (8.3 seconds), while Action Execution accounts for about 9\% (4.6 seconds). 
In summary, the goal-based navigation efficiency is primarily constrained by interpretation and decision-making overhead, and the action execution requires a relatively shorter time.

\subsection{Effectiveness of Exploratory Goal Navigator Task}

The 360-degree exploratory scanning task was evaluated across three environments.
Table~\ref{tab:scan_eval} shows the three attempts for each environment.
Column 3 stands for the number of rotations before a scan was considered complete.
On average, the time cost for one completed scan is 41.1 seconds, which is less than one minute. Additionally, each rotation only took 4.6 seconds, which indicates the framework \tool{} can finish the task in a timely manner.
However, the number of rotations varied widely across environments, ranging from as few as 2 to as many as 12. This variability indicates that the LLM struggled to consistently determine the appropriate stopping condition during exploratory tasks.

\begin{table}[t]
\centering
\caption{Evaluation results for exploratory scanning task.}
\label{tab:scan_eval}
\renewcommand{\arraystretch}{1.1}
\setlength{\tabcolsep}{2.5pt} 
\begin{tabular}{|c|c|c|c|c|c|}
\hline
Env. & Att. & Rot. & 
\begin{tabular}[c]{@{}c@{}}Time/\\ Rot. (sec.)\end{tabular} & 
\begin{tabular}[c]{@{}c@{}}Decision\\ (sec.)\end{tabular} & 
\begin{tabular}[c]{@{}c@{}}Total/\\ Att. (sec.)\end{tabular} \\ \hline

\multirow{3}{*}{Highway} 
 & 1 &  4 & 3.65 &  4.74 & 19.35 \\ \cline{2-6} 
 & 2 &  5 & 3.75 &  4.92 & 23.64 \\ \cline{2-6} 
 & 3 &  5 & 3.42 &  3.33 & 20.44 \\ \hline

\multirow{3}{*}{Country} 
 & 1 & 10 & 5.21 & 12.67 & 64.76 \\ \cline{2-6} 
 & 2 &  8 & 5.31 &  5.92 & 48.42 \\ \cline{2-6} 
 & 3 &  2 & 2.87 &  5.74 & 14.02 \\ \hline

\multirow{3}{*}{Ship} 
 & 1 &  9 & 4.90 &  6.66 & 50.77 \\ \cline{2-6} 
 & 2 & 10 & 5.12 &  8.39 & 59.57 \\ \cline{2-6} 
 & 3 & 12 & 5.32 & 11.15 & 75.04 \\ \hline
\end{tabular}
\vspace{-0.3cm}
\end{table}

%% file: content/discussion.tex
\section{Discussion}

In this section, we will discuss the usability of adopting the LLM solution on
navigation tasks based on the evaluation results of \tool{}, and identify the shortcomings that need to be addressed in future work.

\subsection{Usability Discussion}

\textbf{Action Navigator.}
This component consistently demonstrates high accuracy, correctly analyzing 21 out of 21 user queries' intentions and successfully mapping them to the pre-defined actions.
In addition, the low overhead (approximately 0.72 seconds) shows strong potential in handling the live navigation request.
Together, these results highlight Action Navigator's reliable usability and minimal impact on overall performance.

\textbf{Comprehensive Interpreter.}
The Comprehensive Interpreter enables accurate perception of the navigation context, supporting the AI model in making reliable decisions.
As a result, \tool{} successfully completed 16 out of 18 direct goal-oriented navigation tasks.
However, this component also introduces substantial latency, which accounts for approximately 72\% of the execution time per navigation turn, with visual interpretation contributing 40\% and textual description 32\%.
This indicates that both sub-processes are major bottlenecks and need to be optimized to enhance overall system performance.
In summary, while the Comprehensive Interpreter is suitable for time-insensitive tasks such as functional testing, its high latency limits its usability for real-time applications like live support.

\textbf{Decision Voter.}  
The Decision Voter integrates multiple LLMs to enable robust decision-making, achieving a high success rate of 89\% (16 out of 18 attempts). The average overhead is 8.3 seconds, primarily due to sending and waiting for responses from three separate AI models.
Therefore, the average time cost per model is 2.76 seconds, which is more acceptable. 
However, during exploratory scanning tasks, the number of rotations varied widely from 2 to 12, indicating the difficulty that the large language models have in consistently determining the scan termination conditions. 
This inconsistency reveals the models' limitations in understanding a sequence of image inputs and a potential need for more complicated context management.
In summary, while Decision Voter demonstrates its effectiveness in handling the offline testing, its usability decreases in real-time navigation support.

\textbf{Overall Usability Assessment.}  
Across environments, the \tool{} framework demonstrates a high success rate ( approximately 89\%) and provides valuable insights into LLM-driven navigation.
However, its overall efficiency is constrained by LLM-based interpretation and decision-making costs, and inconsistent termination behavior further reduces scanning task efficiency. 
In summary, the \tool{} solution is \textit{reasonable for offline testing and evaluation}, but its usability is \textit{limited for real-time or live support applications} without further optimization.

\subsection{Future Research Insight}

To address the usability limitations of the Comprehensive Interpreter, the main goal is to reduce processing time for both text generation and image recognition. Currently, \tool{} relies entirely on remote LLMs for these tasks, leading to longer delays. A promising direction is to decompose the process into subtasks and develop lightweight local AI models to handle them more efficiently. The key research challenge lies in defining these subtasks and building fast, accurate local models. This aligns with recent trends in edge-based, low-latency LLM design~\cite{pollini2025reducinglatencyllmbasednatural, zheng2025reviewonedgellms}.

For the Decision Voter, the main usability issues stem from the sequential communication with multiple LLM agents and the lack of effective context management.
The first issue—latency due to sequential processing—can be mitigated by adopting parallel inference and reducing token usage, similar to the strategies proposed in Plurals~\cite{Plurals2025CHI} and LiveMind~\cite{chen2024livemindlowlatencylargelanguage}.
The second issue requires designing efficient mechanisms to manage and synchronize context across multiple agents.

%% file: content/conclusion.tex
\section{conclusion}

In this paper, we presented \tool{}, a large language model-powered framework for automating task-oriented navigation in VR environments. 
It integrates virtual scene interpretation, navigation query classification, multi-agent decision voting, and control mapping, which simulate navigation actions.
\tool{} demonstrated strong effectiveness in supporting Action Navigator and Direct Goal-oriented tasks, achieving an 89\% success rate across three VR environments. However, the observed overhead from interpretation and decision-making significantly reduces the navigation usability. 
We conclude with a discussion on these limitations and propose directions for future improvements.